# Simple illustrations of range-dependence and 3-D effects by normal-mode sound propagation modelling

Sven Ivansson[1]

[1] Swedish Defence Research Agency, SE-16490 Stockholm, Sweden
Contact email: sven.ivansson@foi.se

## Abstract

As is well known, the sound-speed profile has significant effects on underwater acoustic sound propagation. These effects can be quantified by normal-mode models, for example. The basic case is a laterally homogeneous medium, for which the sound speed and the density depend on depth only and not on horizontal position. Effects of horizontal medium-parameter variation can be quantified by coupled-mode models, with coupling between mode expansions for laterally homogeneous parts of the medium. In the present paper, these effects are illustrated for media with a particularly simple horizontal parameter variation such that mode shapes do not vary with horizontal position. The modal wavenumbers depend on horizontal position, however. At a vertical interface between regions with laterally homogeneous medium parameters, each mode is reflected as well as transmitted. For the media considered, reflection and transmission coefficients can be computed separately for each mode without mode coupling, and this is done recursively, stepping through the laterally homogeneous regions in appropriate order. Continuous-wave as well as pulse computations are used to illustrate frequency-dependent reflections from ridges and diffraction into shadow zones behind islands, for example.

## 1 Introduction

Time-harmonic propagation of sound waves in a fluid medium, with a water column on top, is governed by the Helmholtz equation. Assuming that the density $\rho$ and the sound speed $c$ do not depend on horizontal position, only on depth $z$ below the water surface, and that the field is driven by a symmetric point source, the Helmholtz equation can be solved by separation of variables. The depth-separated Helmholtz equation

$$\rho(z) \frac{d}{dz}\left(\frac{1}{\rho(z)} \frac{dZ(z)}{dz}\right) + \left(\frac{\omega^2}{c^2(z)} - k^2\right) Z(z) = 0 \qquad (1)$$

appears, where $Z(z)$ is a pressure mode-shape function and $k$ is a modal wavenumber [1]. A dependence on time $t$, given by the typically omitted factor $\exp(-i\omega t)$ is assumed, where $\omega$ is the angular frequency. The accompanying differential equation for the range-dependence factor is solved by zeroth-order Hankel functions. Equation (1) is to be solved together with appropriate boundary conditions at the surface ($z = 0$) and at the bottom. A pressure-release surface, $Z(0) = 0$, is typically assumed. Non-trivial solutions only appear for a discrete set of modal wavenumbers $k = k_1, k_2, \ldots$ with corresponding mode functions $Z_1(z), Z_2(z), \ldots$

      A modified medium is now considered, with sound-speed and density functions changed to $R\rho(z)$ and $(1/c^2(z)+S)^{-1/2}$, respectively, where $R > 0$ and $S$ are constant

numbers. Equation (1) immediately shows, cf. Section 7.1.2 in [2], that this medium has the *same* mode functions $Z_m(z)$, m=1,2,.., as the original medium, but the corresponding modal wavenumbers are changed from $k_m$ to $(k_m^2 + \omega^2 S)^{1/2}$, m=1,2,...

In the present paper, range-dependence and 3-D effects are illustrated using media built up by regions differing from each other only by different *R* and *S* parameter values for the density and sound-speed profiles, respectively. Apparently, it is sufficient to resolve the mode structure for the underlying reference medium with *R* = 1 and *S* = 0. A cylindrically symmetric 2-D case is considered in Sec. 2. Sections 3 and 4 deal with 2.5-D cases, for which the medium parameters are independent of one of three spatial coordinates: the Cartesian horizontal coordinate *y* in Sec. 3 and the cylindrical horizontal coordinate *r* in Sec. 4. A symmetric point source is assumed all the time. The difference between Secs. 2 and 4 is that the source is at *r* = 0 and *r* ≠ 0, respectively.

Details of the used computational methods, as well as references to alternative ones, appear in [3]. In essence, reflection and transmission matrices for the modes are computed at the vertical interfaces separating the different medium regions. These matrices are computed recursively, adapting invariant embedding techniques from [4] and [5]. For the simple medium types considered, the reflection matrices are actually diagonal, implying that the different modes can be handled separately. Moreover, a subsequent recursion with transmission matrices can be avoided by determination of a scaling factor for stored modal amplitudes.

## 2 Medium with cylindrical symmetry around the source (2-D): reef example

Cylindrical coordinates $(r,\phi,z)$ are introduced, where $r = (x^2+y^2)^{1/2}$ is radius, i.e., horizontal range, and $\phi$ is azimuth in the horizontal plane. Here, *x* and *y* are Cartesian coordinates in the horizontal plane. For a cylindrically symmetric medium, where $\rho$ and *c* are independent of $\phi$, the Helmholtz equation for the acoustic pressure *p* takes the form

$$\frac{\rho(r,z)}{r}\frac{\partial}{\partial r}\left(\frac{r}{\rho(r,z)}\frac{\partial p}{\partial r}\right) + \rho(r,z)\frac{\partial}{\partial z}\left(\frac{1}{\rho(r,z)}\frac{\partial p}{\partial z}\right) + \frac{1}{r^2}\frac{\partial^2 p}{\partial \phi^2} + \left(\frac{\omega^2}{c^2(r,z)}\right)p = \left(\frac{\omega^2}{c^2(r,z)}\right)M\delta_s . \qquad (2)$$

In the right-hand side, $\delta_s$ is the 3-D Dirac function centred at an assumed symmetric point source with moment-tensor strength *M*.

In the present Sec. 2, the point source is assumed to be at depth $z_s$ on the *z* axis (*r* = 0). Hence, $p = p(r,z)$ is independent of $\phi$. The left part of Fig. 1 illustrates a medium with *N*+1 laterally homogeneous segments, each of which has its own *R* and *S* parameters, as introduced in Sec. 1. The right part shows a view of the horizontal plane for *N* = 2.

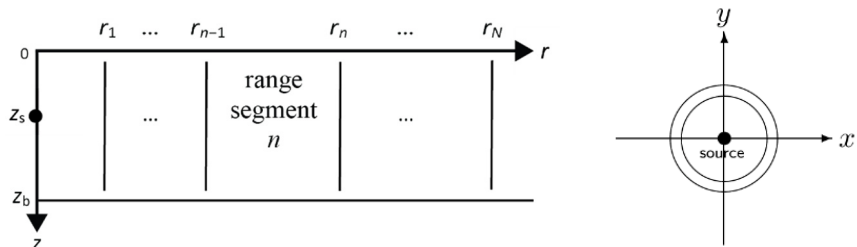

Figure 1: This medium, bounded at depth at z = $z_b$, is cylindrically symmetry around the source.

Figure 2 shows two sound-speed profiles down to a depth of 120 m. The black curve, considered to be the reference medium with $S = 0$, concerns a 100 m deep water column with two underlying 10 m thick sediment layers. The upper sediment layer has sound speed 1500 m/s, density 1500 kg/m$^3$, and absorption 1.0 dB/wavelength. For the lower one, the sound speed is 1600 m/s, the density is 1900 kg/m$^3$, and the absorption is 0.8 dB/wavelength. Below the sediment layers, there is a homogeneous half-space with density 2400 kg/m$^3$. Its wave-speed and absorption data are 3000 m/s and 0.1 dB/wavelength, respectively.

For the grey curve in Fig. 2, the $S$ parameter value is selected to provide a sound speed increase from 1454 to 1800 m/s and an absorption of 0.87 dB/wavelength at depth $z = 10$ m. These values correspond to coarse sand [6]. With $R = 2.2$, a sand density value of 2200 kg/m$^3$ is matched for depths down to $z = 100$ m.

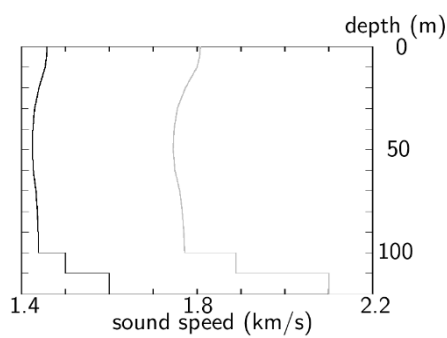

Figure 2: Sound-speed profiles in the water (black curve) and in the sand (grey curve). Two sediment layers below the water depth 100 m are also included.

Some propagation loss (PL) results for a source at depth $z_s = 10$ m follow in Figs. 3 and 4. There are two panels in each figure, the left one is for 100 Hz and the right one is for 500 Hz. Figure 3 concerns a laterally homogeneous reference medium, with $R = 1$ and $S = 0$ throughout, for which the grey curve in Fig. 2 is not involved. Reflections from the water/sediment interface are clearly seen at 500 Hz for the lower grazing angles, while much of the steeply propagating energy is lost into the bottom.

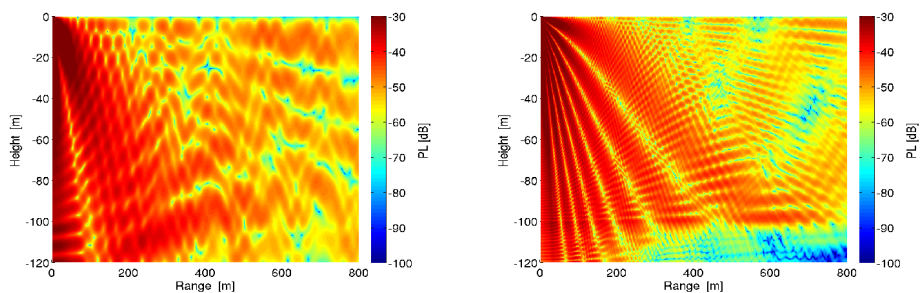

Figure 3: Propagation loss (PL) in the range,depth $(r,z)$ plane for the laterally homogeneous reference medium. The source is at depth $z_s = 10$ m. *Left*: 100 Hz. *Right*: 500 Hz.

With a ring-shaped sand reef between $r = 250$ m and $r = 350$ m, cf. the right part of Fig. 1, Fig. 4 indicates significant loss at propagation through the sand. As expected, the loss increases with frequency. Some energy is reflected back from the vertical water/sand interface, causing increased energy levels for $r < 250$ m, compared to Fig. 3.

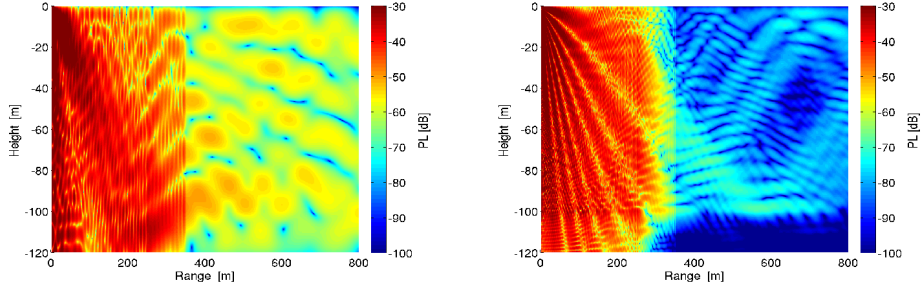

Figure 4: As Fig. 3 but for the medium with the sand reef at 250 m < *r* < 350 m.

## 3 Medium invariant in the Cartesian y coordinate (2.5-D): ridge example

When $\rho$ and $c$ are independent of the Cartesian horizontal coordinate $y$, the Helmholtz equation for the acoustic pressure $p$ takes the form

$$\rho(x,z)\frac{\partial}{\partial x}\left(\frac{1}{\rho(x,z)}\frac{\partial p}{\partial x}\right) + \rho(x,z)\frac{\partial}{\partial z}\left(\frac{1}{\rho(x,z)}\frac{\partial p}{\partial z}\right) + \frac{\partial^2 p}{\partial y^2} + \left(\frac{\omega^2}{c^2(x,z)}\right)p = \left(\frac{\omega^2}{c^2(x,z)}\right)M\delta_s . \quad (3)$$

As in Eq. (2), $\delta_s$ is the 3-D Dirac function centred at an assumed symmetric point source with moment-tensor strength *M*. Fourier transformation of Eq. (3) shows that

$$p(x,y,z) = \int_{-\infty}^{+\infty} \hat{p}(x,z;\kappa)\, e^{i\kappa y}\, d\kappa , \quad (4)$$

where, for each $\kappa$, $\hat{p}(x,z;\kappa) = (2\pi)^{-1}\int_{-\infty}^{+\infty} p(x,y,z)\, e^{-i\kappa y}\, dy$ fulfils the 2-D Helmholtz equation

$$\rho(x,z)\frac{\partial}{\partial x}\left(\frac{1}{\rho(x,z)}\frac{\partial \hat{p}}{\partial x}\right) + \rho(x,z)\frac{\partial}{\partial z}\left(\frac{1}{\rho(x,z)}\frac{\partial \hat{p}}{\partial z}\right) + \left(\frac{\omega^2}{c^2(x,z)} - \kappa^2\right)\hat{p} = \left(\frac{\omega^2}{c^2(x,z)}\right)\frac{M\delta(x)\delta(z-z_s)}{2\pi} . \quad (5)$$

Here, $\delta$ is the 1-D Dirac function and the symmetric point source is still assumed to be on the *z* axis at $(x,y,z) = (0m,0m,z_s)$. The problem is thereby reduced to numerical integration of the solutions to a number of $\kappa$–dependent 2-D problems. These 2-D problems are of a type related to the one in Sec. 2, and adaptive integration is a convenient method for recovering *p* according to Eq. (4). A resonance phenomenon appears when $\kappa$ is equal to a modal wavenumber for a specific laterally homogeneous region, cf. Sec. 4.4 in [7]. Hence, for a medium with vanishing or very small absorption, the integration path must in general be displaced from the real axis. Details of the computation method with reflection matrices are given in [3].

The left part of Fig. 5 illustrates a medium with *N*+1 laterally homogeneous segments, each of which has its own *R* and *S* parameters, as introduced in Sec. 1. The right part shows a view of the horizontal plane for *N* = 2.

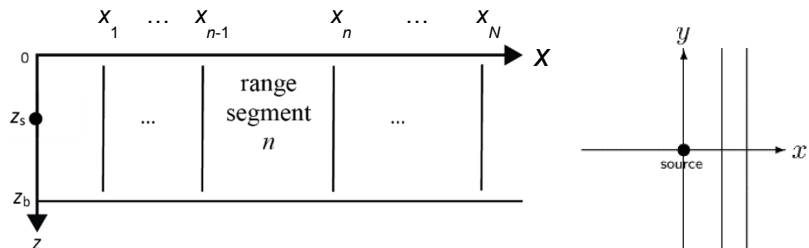

Figure 5: The medium parameters are here independent of the *y* coordinate.

Figure 6 shows PL results at depth 90 m when the source is at (*x*,*y*,*z*) = (0m,0m,10m) and there is a sand ridge between *x* = 250 m and *x* = 350 m. As in Fig. 4, significant frequency-dependent loss is suffered at propagation through the sand. At large incidence angles (large *y*), however, when a critical angle is passed, most of the energy is reflected back. As a result, clear distortions of the near-field Lloyd mirror circles (caused by interference with reflections from the surface [1]) can be observed at large *y*.

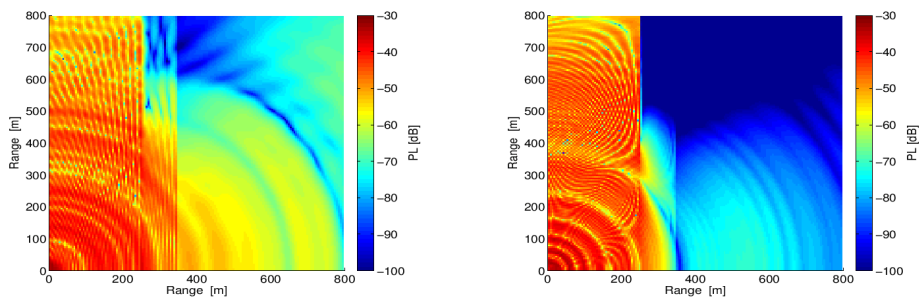

Figure 6: PL in a horizontal (*x*,*y*) plane for a medium with a sand ridge at 250 m < *x* < 350 m. The source and receiver are at depths 10 and 90 m, respectively. *Left*: 100 Hz. *Right*: 500 Hz.

The time signal from a transient source pulse can be modelled by Fourier synthesis. Figure 7 shows results for the laterally homogeneous reference medium with *R* = 1 and *S* = 0 throughout. A broad-band source pulse is assumed, centred at 100 and 500 Hz for the left and right panels, respectively. The source is still at (*x*,*y*,*z*) = (0m,0m,10m), but there are now fifteen receivers at (*x*,*y*,*z*) = (150m,*y*,90m) for *y* = 50, 100, .., 750 m. Two groups of arrivals appear: an initial one which has passed only once between the source and receiver depths, and a latter one with an additional down-up cycle between the source and receiver depths. Reduced time, $t - y/(1480\text{m/s})$ is used to achieve concentrated plots.

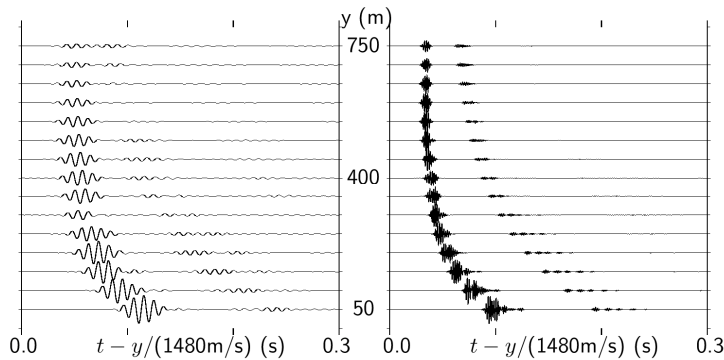

Figure 7: Time traces for propagation in the laterally homogeneous reference medium. The source is at (*x*,*y*,*z*) = (0m,0m,10m), and there are 15 receivers at (*x*,*y*,*z*) = (0m,*y*,90m) for *y* = 50, 100, .., 750 m. The centre frequency of the broad-band source pulse is 100 Hz in the left panel and 500 Hz in the right panel.

Figure 8 shows corresponding time traces for the medium with a sand ridge between $x = 250$ m and $x = 350$ m. Two additional arrival groups, involving reflections from the vertical water/sand interface, are clearly seen in the right panel. The strong initial one has passed only once between the source and receiver depths, and the latter one has completed an additional down-up cycle between the source and receiver depths. As expected, the water/sand reflections are much weaker for the low-frequency pulse.

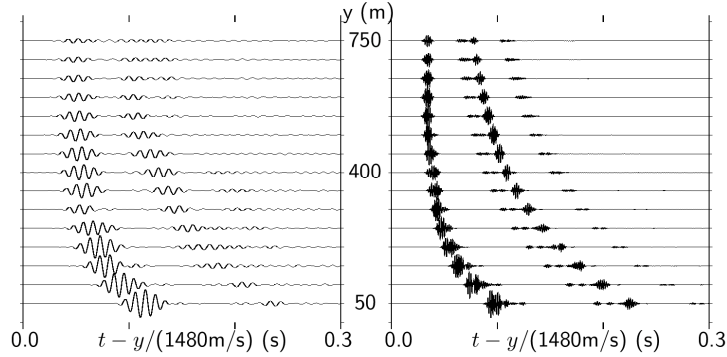

Figure 8: As Fig. 7 but for the medium with the sand ridge at 250 m < $x$ < 350 m.

Figure 9 is similar to Fig. 8, but a transition zone is now introduced between $x = 200$ m and $x = 250$ m, where the medium parameters change gradually from water to sand. The transition zone is modelled by 50 laterally homogeneous strips, each of $x$ width 1 m. For the receivers at $(x,y,z) = (150\text{m},y,90\text{m})$ with $y < 600$ m, the water/sand reflections for the pulse centred at 500 Hz are now much weaker than in Fig. 8. At the larger receiver-$y$, however, these reflections are still strong. For the corresponding larger incidence angles, the energy turns back towards the water before suffering significant losses in the transition zone. In effect, a kind of "horizontal refraction" appears.

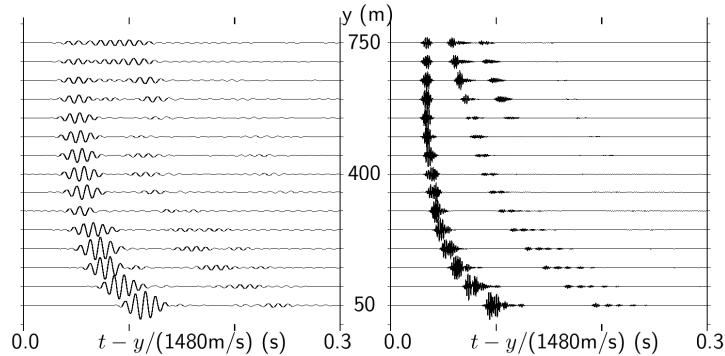

Figure 9: As Fig. 8 but with a gradual transition from water at $x = 200$ m to sand at $x = 250$ m.

## 4 Cylindrically symmetric medium, displaced source (2.5-D): island example

The point source is now assumed to be at $(x,y,z) = (x_s,0\text{m},z_s)$, in cylindrical coordinates $(r,\phi,z) = (r_s,0,z_s)$. Here, $x_s = r_s > 0$. When $c$ and $\rho$ are independent of the cylindrical coordinate $\phi$, Fourier transformation of Eq. (2) shows that

$$p(r,\phi,z) = \sum_{\nu=0}^{\infty} \hat{p}(r,z;\nu)\,(2-\delta_{\nu 0})\cos(\nu\phi)\,, \qquad (6)$$

where $\hat{p}(r,z;\nu) = \frac{1}{2\pi}\int_0^{2\pi} p(r,\phi,z)\cos(\nu\phi)\,\mathrm{d}\phi = \frac{1}{\pi}\int_0^{\pi} p(r,\phi,z)\cos(\nu\phi)\,\mathrm{d}\phi$ and $\delta_{\nu 0}$ is the Kronecker delta. For each $\nu$, $\hat{p}$ fulfils the 2-D Helmholtz equation

$$\frac{\rho(r,z)}{r}\frac{\partial}{\partial r}\left(\frac{r}{\rho(r,z)}\frac{\partial \hat{p}}{\partial r}\right) + \rho(r,z)\frac{\partial}{\partial z}\left(\frac{1}{\rho(r,z)}\frac{\partial \hat{p}}{\partial z}\right) + \left(\frac{\omega^2}{c^2(r,z)} - \frac{v^2}{r^2}\right)\hat{p}$$
$$= \left(\frac{\omega^2}{c^2(r,z)}\right)\frac{M\delta(r-r_s)\delta(z-z_s)}{2\pi r}. \qquad (7)$$

The problem is thereby reduced to summation of the solutions to a number of $v$–dependent 2-D problems that are of a type related to the one in Sec. 2. Again, details are given in [3].

The left part of Fig. 10 illustrates a medium with $N+1$ laterally homogeneous segments, each of which has its own $R$ and $S$ parameters, as introduced in Sec. 1. Since the segment numbers increase from the source, assumed to be in the outermost segment, the interface ranges $r_n$ now decrease when $n$ increases. The right part shows a view of the horizontal plane for $N = 1$.

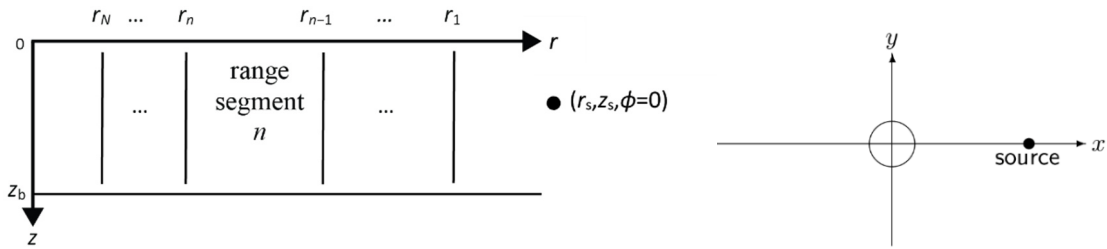

Figure10: As in Fig. 1, the medium is cylindrically symmetric. The source is now displaced from the vertical symmetry axis, however.

As in all previous result figures, the source depth $z_s$ is 10 m in Figs. 11-17. In contrast to the previous figures, however, the source is not at $(x,y) = (0\text{m},0\text{m})$ but at $(x,y) = (300\text{m},0\text{m})$, i.e., $x_s = r_s = 300$ m. Figures 11-12 and 13-14 show PL results in the range,depth plane for $y = 0$ m and in the horizontal plane at receiver depth 90 m, respectively. There is a disk-shaped island at $r < 50$ m: a sand island in Figs. 11 and 13, and a rigid island in Figs. 12 and 14.

Figure 11 can be compared to the 2-D results in Fig. 4. In both cases, the energy propagates for a horizontal range of 250 m before meeting sand of horizontal width 100 m. Diffraction effects cause somewhat higher energy levels behind the sand in Fig. 11 than in Fig. 4.

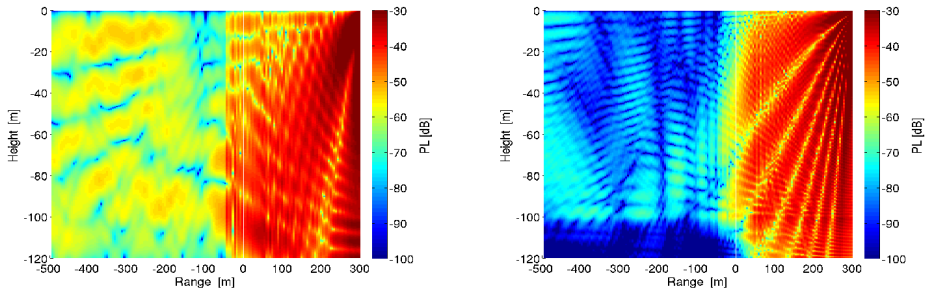

Figure11: PL in the range,depth (*x,z*) plane at *y* = 0 m for the medium with a disk-shaped sand island at *r* < 50 m. The source is at (*x,y,z*) = (300m,0m,10m). *Left*: 100 Hz. *Right*: 500 Hz.

The diffraction effects are very clear in Fig. 12, with a rigid island at $r < 50$ m. This time, there is a geometric shadow zone behind the island, and a 2-D computation would predict vanishing energy levels (infinite PL) there. As expected, the diffraction into the shadow zone decreases when the frequency is increased.

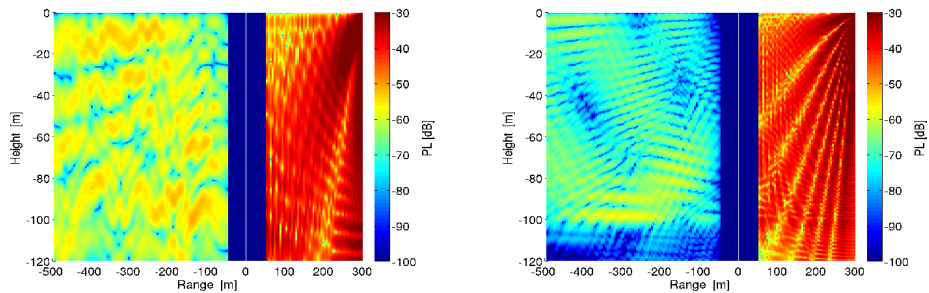

Figure 12: As Fig. 11 but for a rigid island at *r* < 50 m.

The horizontal view in Fig. 13, at the receiver depth 90 m, shows clearly how energy is successively lost at propagation through the sand island. As in Fig. 6, Lloyd mirror effects show up as circles centred at the source.

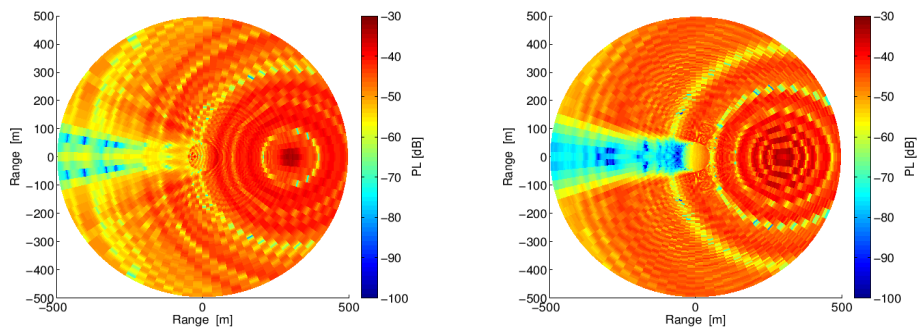

Figure 13: PL in the horizontal (*x*,*y*) plane at receiver depth 90 m for the medium with a disk-shaped sand island at *r* < 50 m. The source is at (*x*,*y*,*z*) = (300m,0m,10m). *Left*: 100 Hz. *Right*: 500 Hz.

At 500 Hz, the shadow behind the island is less developed in Fig. 14 than in Fig. 13. The significant anelastic high-frequency loss in the sand is of course avoided when the island is rigid. Simplified N×2D computations, with 2-D computations as in Sec. 2 for a number of vertical planes through the source, would erroneously predict an impenetrable shadow zone behind a rigid island.

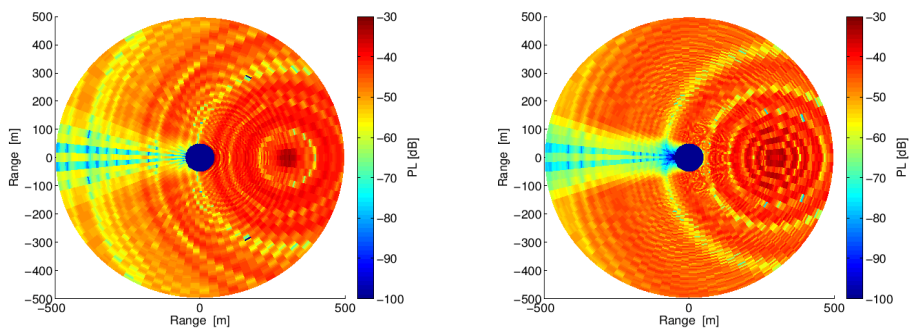

Figure 14: As Fig. 13 but for a rigid island at *r* < 50 m.

Figures 15-17 show time signals from a transient source pulse, obtained by Fourier synthesis. Figure 15 concerns the laterally homogeneous reference medium with *R* = 1 and *S* = 0 throughout. The major difference to Fig. 7 is that the receivers, still at depth 90 m, are now located in a semi-circle at *r* = 150 m: from (*x*,*y*) = (150m,0m), i.e., $\phi = 0°$, to (*x*,*y*) = (−150m,0m), i.e., $\phi = 180°$. As in Fig. 7, two groups of arrivals appear: an initial

one which has passed only once between the source and receiver depths, and a latter one with an additional down-up cycle between the source and receiver depths.

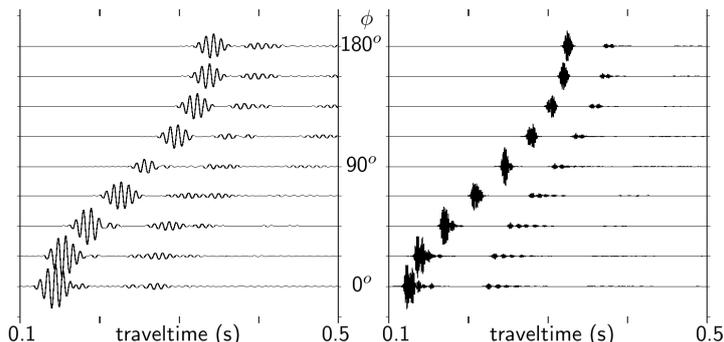

Figure15: Time traces for propagation in the laterally homogeneous reference medium. The source is at (*x*,*y*,*z*) = (300m,0m,10m). There are nine receivers at (*r*,$\phi$,*z*) = (150m,$\phi$,90m): $\phi$ = 0º,22.5º,..,180º. The centre frequency of the broad-band source pulse is 100 Hz in the left panel and 500 Hz in the right panel.

An island at *r* < 50 m is included in Figs. 16 (the sand island) and 17 (the rigid island). As expected, reflections from the island appear, that are strongest for the rigid case. Arrivals at $\phi$ close to 180º are shadowed by the island. Weaker diffracted arrivals can be seen, however, particularly in Fig. 17 where there is no high-frequency loss in anelastic sand.

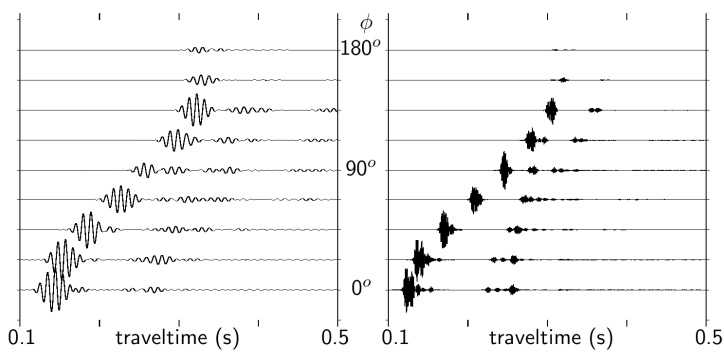

Figure16: As Fig. 15 but for the medium with the sand island at *r* < 50 m.

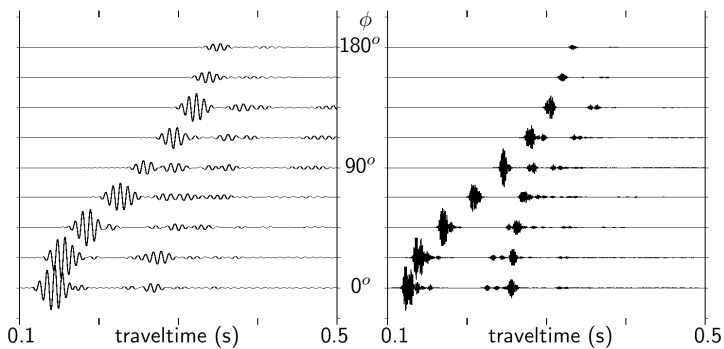

Figure17: As Figs. 15 and 16 but for the medium with the rigid island at *r* < 50 m.

## 5 Conclusions

As detailed in [3], sound propagation in the following three types of laterally varying fluid media can be modelled using recursive computation of reflection and transmission matrices for normal modes:

- cylindrically symmetric media with the source on the vertical symmetry axis (2-D computations)
- *y*-invariant media (2.5-D computations)
- cylindrically symmetric media with the source displaced from the vertical symmetry axis (2.5-D computations).

Some effects of lateral variation are illustrated for these three types of media in Secs. 2, 3, and 4, respectively. In each case, the medium is built up by laterally homogeneous regions. The lateral variation among these regions is of a particular type, such that the mode shapes are the same while the lateral variation manifests itself by different modal wavenumbers. As a consequence, the coupling matrices between modes in different regions are diagonal, and the modes can be handled separately.

Further work of interest includes implementation of more general cases with non-diagonal coupling matrices, extension to the fluid-solid case, and approximations for full 3-D cases.